\documentclass{llncs}

\usepackage{makeidx}
\usepackage{color,url,times}
\usepackage{graphicx,color}
\usepackage{tikz}
\usetikzlibrary{arrows,shapes}
\usepackage{floatflt,enumitem}
\usepackage{mdframed,framed}
\usepackage{amssymb}
\usepackage{amsmath}
\usepackage{makeidx}  
\usepackage{url}
\usepackage{mdframed,framed}

\newcommand{\ass}{\texttt{:=}}     
\newcommand{\inc}{~~~~\= \+ \kill}    
\newcommand{\dec}{\- \kill}         

\newcommand{\reserved}[1]{\textbf{\underline{#1}}} 
\newcommand{\semicol}{;}                  

\newcommand{\IF}{\reserved{if}}

\newcommand{\DO}{\reserved{do}}
\newcommand{\OD}{\reserved{end~do}}
\newcommand{\THEN}{\reserved{then}}
\newcommand{\ELSE}{\reserved{else}}
\newcommand{\WHILE}{\reserved{while}}

\begin{document}

\mainmatter
\pagestyle{headings}

\title{Symbolic Computation and Automated Reasoning \\for\\ Program Analysis}
\titlerunning{SYMCAR}

\author{Laura Kov\'acs}

\authorrunning{L.~Kov\'acs}

\institute{Chalmers University of Technology \\and\\
TU Wien}

\maketitle

\vspace*{-1em}
\begin{abstract}
This talk describes 
how a combination of symbolic computation techniques with first-order theorem proving can be used for solving some challenges of automating program analysis, in particular for generating and proving properties about the logically complex parts of software.
The talk will first present how computer algebra methods, such as Gr\"obner basis computation, quantifier elimination and algebraic recurrence solving, help us in 
inferring properties of program loops with non-trivial
arithmetic. Typical properties inferred by our work are loop
invariants and expressions bounding the number of loop iterations.
The talk will then describe our work to generate first-order
properties of programs with unbounded data structures, such as arrays.
 For doing so, we use saturation-based first-order theorem proving 
and extend first-order provers with support for  program analysis. 
Since program analysis requires reasoning in the combination of first-order theories of data structures, 
the talk also discusses new features in first-order theorem proving,
such as inductive reasoning and built-in boolean sort. 
These extensions allow us to express program properties 
directly in first-order logic and hence use further first-order theorem provers to reason about program properties.

\end{abstract}


\section{Introduction}

Individuals, organizations, industries, and nations are increasingly depending on software and systems using
software. This software is large and complex and integrated in a continuously changing complex environment.
New languages, libraries and tools increase productivity of programmers, creating even more software, but the
reliability, safety and security of the software that they produce is still low. We are getting used to the fact
that computer systems are error-prone and insecure. Software errors cost world economies billions of euros.
They may even result in loss of human lives, for example by causing airplane or car crashes, or malfunctioning
medical equipment. 
To improve software and methods of software development one can use a variety of
approaches, including automated software verification and static analysis of programs.
 
The successful development and application of powerful verification
tools such as model checkers~\cite{ClarkeEmerson:LogicPrograms:ModelChecking:1982,Sifakis:TCS:ModelChecking:1982}, 
static program analyzers~\cite{Cousot77}, symbolic computation algorithms~\cite{Buchberger06},
decision procedures for common data structures~\cite{Nelson66},
as well as theorem provers for first- and higher-order
logic~\cite{HandbookAR} opened new perspectives for the automated verification of software systems.
In particular, increasingly common use of concurrency in the new generation
of computer systems has motivated the integration of established
reasoning-based methods, such as satisfiability modulo
theory (SMT) solvers and first-order  theorem
provers, with complimentary techniques such as software testing~\cite{Testing04}.
This kind of integration has however imposed new requirements on
verification tools, such as inductive
reasoning~\cite{McMillan08,Kovacs08d}, 
interpolation~\cite{McMillanTacas06}, proof
generation~\cite{ProofsZ3}, and  non-linear arithmetic symbolic 
computations~\cite{MouraCADE13}. 
Verification methods combining symbolic computation and 
automated reasoning are  therefore of critical importance 
for improving software  reliability. 

In this talk we address this challenge by automatic program
analysis. Program analysis aims to discover program
properties preventing programmers from introducing errors while making
software changes and can drastically
cut the time needed for program development, making thus a crucial step
to automated~verification. The work presented in this
talk targets the combination of symbolic computation techniques from algorithmic
combinatorics and computer algebra with first-order theorem proving and static analysis
of programs. 
 We rely on our
recent {\it symbol elimination method}~\cite{Kovacs08d}. Although the symbol elimination terminology has been introduced only recently
by us,
we argue that symbol elimination can be viewed
as a general framework for program analysis. That is,
various techniques used in software analysis and verification, such as Gr\"obner
basis computation or quantifier elimination,  can be seen as
application of symbol elimination to safety verification of programs.

In  a nutshell, symbol elimination  is based on the following ideas. Suppose we have a
program $P$ with a set of
 variables $V$. The set $V$ defines the language of $P$. We extend the
language $P$ to a richer language $P_0$ by adding functions and
predicates, such as loop counters.
After that,
we automatically generate a set $\Pi$ of first-order properties of the program in the
extended language $P_0$, by using techniques from symbolic computation
and theorem proving. These properties are valid properties of the
program, however they use the extended language $P_0$.
At a last step of symbol elimination we derive from $\Pi$ program
properties in the original language $P$, thus ``eliminating'' the
symbols in $P_0\setminus P$.

The work presented in this talk describes symbol elimination 
in the combination of first-order 
theorem proving  and symbolic computation. 
Such a combination  requires the
development of new reasoning methods based on superposition 
first-order theorem proving~\cite{CAV13}, Gr\"obner basis computation~\cite{Buchberger06},
and quantifier elimination~\cite{Collins75}.   
We propose symbol elimination as
a powerful tool for program analysis, in particular for generating program properties,
such as loop invariants and Craig interpolants.  Such properties express conditions to hold at intermediate program
locations and are used to prove the absence of program errors, hence
they 
are very important for improving automation of program analysis.

Since program analysis requires reasoning in the combination of first-order theories of data structures, 
the talk also presents new features in first-order theorem proving,
such as inductive reasoning and built-in boolean sort. 
These extensions allow us to express program properties 
directly in first-order logic and hence use further first-order theorem provers to reason about program properties. 

The algorithms described in this talk are supported by the development of the world-leading theorem prover
  Vampire~\cite{CAV13}, 
and its extension 
to support program analysis. Thanks to the full automation and tool support of our
work, researchers and  software engineers/developers are able
to use our results in their work, without the need to
become experts in first-order theorem proving and symbolic computation.

The work presented here is structured as follows. We first
describe the use of symbol elimination in symbolic computation for
generating polynomial program properties (Section~\ref{sec:sesc}). We
then extend symbol elimination to its use in first-order theorem
proving and present how arbitrarly quantified program properties can
be inferred using symbol elimination (Section~\ref{sec:setp})

\section{Motivating Example}

\begin{floatingfigure}[r]{80mm}
\vspace*{-.5em}
\begin{framed}
\vspace*{-1.5em}
\begin{tabbing}
$a$ \ass\ $0$; $b$ \ass\ $0$; $c$ \ass\ $0$; $s$ \ass $0$;\\
\WHILE\ ($a < n$) \DO \\ \inc
  \IF\ $A[a] > 0$ \\ \inc
    \THEN\ \=\+ $B[b]$ \ass\ $A[a]+h(b)$\semicol $b$ \ass\ $b+1$\semicol \\ \dec
    \ELSE\ \=\+ $C[c]$ \ass\ $A[a]$\semicol $c$ \ass\ $c+1$\semicol \\ \dec \dec
  $a$ \ass\ $a+1$\semicol \ $s$\ass $s + a*a$\semicol \\ \dec
\OD\\
\reserved{\texttt{assert}}($(\forall p) (0\leq p<b \implies
B[p]-h(p)>0) \ \wedge $\\ 
{\color{white}{\texttt{assert}}(}$\ 6*s=n*(n+1)*(2*n+1)$)
\end{tabbing}
\vspace*{-1.2em}
\caption{Motivating example.\label{fig:partition}}
\end{framed}
\vspace*{-.75em}
\end{floatingfigure}

Let us first motivate the work described in this talk on a small example.
Consider the program  given in Figure~\ref{fig:partition},
written in a C-like syntax. 
The program fills an integer-valued array $B$ 
by the positive values of
a source array $A$ added to the values of a function call $f$, and an
integer-valued array $C$ with the non-positive values of
$A$. In addition, it computes the sum $s$ of squares of the visited positions in $A$. 
A safety assertion, in first-order logic (FOL),  is specified at the end of the loop,
using the {\tt assert} construct.    
The program of Figure~\ref{fig:partition} is
clearly safe as the assertion is satisfied when the loop is exited. However, to prove program
safety we need additional loop properties, i.e. invariants,  that hold at
any loop iteration. 
It is
not hard to derive that  after any iteration $k$ of the loop (assuming
$0\leq k \leq n$), the linear invariant relation $a=b+c$ holds. 
 It is also 
not hard to argue that, upon exiting the loop, the value of $a$ is $n$. 
However, such properties do not give us much information about the
arrays $A$, $B$, $C$ and the integer $s$.  
For proving program safety, we need to derive that each 
$B[0],\dots,B[b-1]$ is the sum of a strictly positive element in $A$
and the value of $f$ at the corresponding position of $B$. We also
need to infer that $s$ stores the sum of squares of the first $n$ non-negative
integers, corresponding to the visited positions in $A$. Formulating
these properties in FOL yields the loop invariant: 
\begin{equation}\label{eq:inv}
\begin{array}{l}
\hskip-.15em (\forall p) (0\leq p<b \implies\\
\qquad  (\exists q) (0\leq q<a\ \wedge\
A[q]>0\ \wedge\ B[p]=A[q]+h(p))  \ \wedge\\
\qquad \ 6*s=a*(a+1)*(2*a+1) )
\end{array}
\end{equation}

The above property  requires quantifier
alternations and polynomial arithmetic and can be used 
to prove the safety assertion of the program.  
This  loop property 
in fact describes much of the intended behavior of the loop and can be
used to analyze 
properties of programs in which this loop is
embedded. 
Generating
 such loop invariants requires however reasoning in full FOL with
 theories, in our example in the first-order 
theory of arrays, polynomial arithmetic and uninterpreted functions.  
Our work addresses this problem and proposes symbol elimination for
automating 
program analysis. 
. 

\section{Symbol Elimination in Symbolic Computation}\label{sec:sesc}
The first part of this talk 
concerns the 
automatic generation of loop invariants over scalar 
variables. This line of research implements 
general idea of symbol elimination by using techniques from  symbolic
computation, as follows.
Given a loop, we first extend the loop language by a new variable $n$,
called the loop counter. Program variables are then considered as
functions of $n$. Next, we apply methods from algorithmic combinatorics
and compute the values of loop variables at arbitrary
loop iterations as functions of $n$. Finally, we eliminate $n$ using
computer algebra algorithms, and derive polynomial 
relations among program variables as loop invariants.

In our work, 
 we identified a certain family of loops, called P-solvable loops (to
 stand for polynomial-solvable)
with sequencing, assignments and conditionals, where test conditions
are ignored~\cite{Kovacs08a}. For these loops, we developed a new algorithm for generating
polynomial loop invariants. Our method uses algorithmic combinatorics
and algebraic techniques, namely solving linear
recurrences with constant coefficients (so-called C-finite recurrences) or hypergeometric terms,
computing algebraic relations among exponential sequences, and eliminating
variables from a system of polynomial equations. More precisely, the key
steps of using symbol elimination in symbolic computation are as
follows. 
Given a P-solvable loop with nested conditionals, we first rewrite
the loop into a collection of P-solvable loops with assignments
only. Next,
polynomial invariants of  all sequences of P-solvable loops with assignments only
are derived. These invariants describe polynomial relations
valid after the first iteration of the P-solvable loop with nested
conditionals, however they might not be valid after an
arbitrary iteration of the P-solvable loop with nested conditionals.
Therefore, from the ideal of polynomial relations after the first
iteration of a P-solvable loop with nested conditionals, we keep only
those polynomial relations that are polynomial invariants of  the
P-solvable loop with nested conditionals.
In the process of deriving polynomial invariants for a (sequence of)
P-solvable loop(s) with assignments only, we proceed as follows.
We introduce a new variable denoting the loop
counter. Next, recurrence equations  over the loop counter are
constructed, describing the behavior of the
loop variables at arbitrary loop iterations. These recurrence
relations are solved, and closed forms of loop variables are computed
as polynomials of the initial values of variables, the loop counter, and some
new variables in the loop counter so that we infer polynomial relations among the new
variables. The loop counter and variables in the loop counter are
then eliminated by Gr\"obner basis computation to derive a finite set
of polynomial identities among the program variables as invariants. From this finite
set any other polynomial identity that is an invariant of the
P-solvable loop with assignments only can be
derived.

\begin{figure}[t]

\begin{tabular}{c||c||c}
$
\left\{\begin{array}{ll}
a^{(k+1)}=&a^{(k)}+1\\
s^{(k+1)}=&s^{(k)}+a^{(k)}*a^{(k)}
\end{array}\right.
$
&
$
\left\{\begin{array}{lll}
a^{(k)}=&a^{(0)}+k\\
s^{(k)}=&s^{(0)}+\frac{k*(k+1)*(2*k+1)}{6}
\end{array}\right.
$
& 
$\begin{array}{l}
6*s^{(k)}=\\
a^{(k)}*(a^{(k)}+1)*(2*a^{(k)}+1)
\end{array}$\\
& & \\
(i) & (ii) & (iii)
\end{tabular}
\caption{Symbol Elimination in Symbolic
  Computation on Figure~\ref{fig:partition}.\label{fig:poly}}
\end{figure}

To illustrate the workflow proposed above, consider 
Figure~\ref{fig:poly}. Figure~\ref{fig:poly}(i) describes 
 the system of recurrence equations corresponding to the
updates over $a$ and $s$ in Figure~\ref{fig:partition}, where  
$s^{(k)}$ and $a^{(k)}$ denote the values of $s$ and $a$ at the $k$th
loop iteration of Figure~\ref{fig:partition}. That is, program variables become
functions of loop iterations $k$.  The closed form
solutions of Figure~\ref{fig:poly}(i) is given in
Figure~\ref{fig:poly}(ii). After substituting the initial values of
$a$ and $s$,  Figure~\ref{fig:poly}(iii) shows a valid polynomial
identity among the values of $a$ and $s$ at any loop iteration $k$.

Our invariant generation method using symbol elimination in symbolic
computation is proved to be complete in~\cite{Kovacs09a}. By completeness we mean 
that our method generates the basis of the polynomial invariant ideal, and hence
any other polynomial invariant of the P-solvable loop can be derived
from the basis of the invariant ideal.
For doing so, we generalised the invariant generation algorithm
of~\cite{Kovacs08a} for P-solvable loops
 by iteratively computing the polynomial invariant ideal of the loop.
We  proved that this generalisation is sound and complete.
That is, our method infers a basis for the polynomial invariant ideal of the P-solvable
loop in a finite number of steps. Our proof relies on showing that the dimensions of the
prime ideals from the minimal decomposition of the ideals generated at an iteration of
our algorithm either remained the same or decreased at the next iteration of the algorithm.
Since dimensions of ideals are positive integers, our algorithm terminates after a finite
number of iterations.


\section{Symbol Elimination in First-Order Theorem Proving}\label{sec:setp}

In the second part of our talk, we describe the use of symbol
elimination in first-order theorem proving.
The method of symbol elimination using a first-order theorem prover
has been introduced in~\cite{Kovacs08d}.
Unlike all previously known techniques, our
method allows one to generate first-order invariants containing alternations of
quantifiers.
The method is based on automatic analysis of the so-called update
predicates of loops. An update predicate for an array  expresses updates
made to the array. We observe that many properties of update predicates can be extracted
automatically from the loop description and loop properties obtained by
other methods such as a simple analysis of counters occurring in the loop, recurrence
solving and quantifier elimination over loop variables.
In the first step of loop analysis we introduce a
new variable $n$ denoting the loop counter, and use the symbolic
computation framework from Section~\ref{sec:sesc} to generate polynomial
invariants over the scalar loop variables. Scalar and array
variables of the loop are considered as functions of $n$ and the
 language of the loop is extended by these new
function symbols.
Further, the loop language is also extended by the update predicates for arrays and
their properties are added to the extended language too. The update
predicates make use of $n$ and
essentially describe
positions at which arrays are updated,
iterations at which the updates occur and the
update values of the arrays.
As a result of this step of symbol elimination,
a new, extended loop language is obtained, and
a collection $\Pi$ of
valid first-order loop properties expressed in the extended loop language is
derived.
Formulas in $\Pi$ cannot be used as loop invariants, since they use symbols not occurring
in the loop, and even symbols whose semantics is described by the loop itself.
These formulas, being valid properties of the loop, have a useful property: all their consequences
are valid loop properties too. The second phase of symbol elimination
therefore tries
to generate logical consequences of $\Pi$ in the original language of the loop. Any such
consequence is also a valid property of the loop, and hence an
invariant of the loop. Logical consequences
of $\Pi$ are generated by running a first-order saturation theorem prover on $\Pi$ in a way that
it tries to eliminate the newly introduced symbols from
the extended loop language.

Since the symbol elimination method was only recently introduced,
its practical power and limitations were not well-understood.
The main obstacle to its experimental evaluation lied in the fact
that all existing first-order theorem provers lacked several features essential for implementing
our procedure for invariant generation. These features included
reasoning with various theories and procedures for eliminating
symbols. In our work we  addressed these limitations and developed new
features in first-order theorem proving, making first-order theorem
provers better suited for applications of program analysis and
verification.

\bigskip

\noindent{\bf{Acknowledgments.}} 
The work described in this talk is based on joint work with
a number of authors, including Tudor Jebelean (RISC-Linz), Evgeny
Kotelnikov and Simon Robillard (Chalmers University of Technology),
and Andrei
Voronkov (The University of Manchester and Chalmers University of
Technology). 

The author acknowledges funding from the ERC Starting Grant 2014
SYMCAR 639270, the Wallenberg Academy Fellowship 2014, the Swedish
VR grant D0497701 and the Austrian research project FWF S11409-N23

\end{document}